\documentclass[twocolumn,twoside,slac_two]{revtex4}
\usepackage{graphicx}
\usepackage{fancyhdr}
\newcommand{\apj}{\mbox{\it Astrophys. J.}}
\newcommand{\apjl}{\mbox{\it Astrophys. J.}}
\newcommand{\aap}{\mbox{\it Astron. Astrophys.}}

\newcommand{\mnras}{\mbox{\it Mon. Not. R. Astron. Soc.}}
\newcommand{\nat}{\mbox{\it Nature}}

\newcommand{\keV}{\rm{\, keV }}
\newcommand{\MeV}{\rm{\, MeV }}
\newcommand{\GeV}{\rm{\, GeV }}

\def\gtsima{$\; \buildrel > \over \sim \;$}
\def\ltsima{$\; \buildrel < \over \sim \;$}

\def\gsim{\lower.5ex\hbox{\gtsima}}
\def\lsim{\lower.5ex\hbox{\ltsima}}

\begin{document}

%Title of paper
\title{The Impact of Fermi on the Study of Gamma-ray Bursts}

% Repeat the \author .. \affiliation  etc. as needed
%
% \affiliation command applies to all authors since the last
% \affiliation command. The \affiliation command should follow the
% other information

\author{Asaf Pe'er}
\affiliation{Harvard-Smithsonian Center for Astrophysics, MS-51, 60
  Garden Street, Cambridge, MA 02138, USA}

\begin{abstract}
  Recent {\em Fermi} results have focused attention on gamma-ray burst's
  (GRB) prompt emission phase, which is rich in phenomenology and
  poorly understood.  The broad band spectra observed by {\em Fermi} does
  not fit into any of the frameworks of existing theoretical
  models. Thus, {\em Fermi} results force new thinking of questions
  that were thought to be solved. I highlight here the basic open
  questions prior to the launch of {\it Fermi}, key {\em Fermi}
  results, and new theoretical ideas that emerged following these
  results. These include: (I) renewed interest in magnetized outflows
  as a way to understand the dynamics and composition; (II) interest
  in photospheric emission, in particular ways to broaden ``Planck''
  spectrum to resemble the observed ``Band'' spectrum; (III) The
  puzzling origin of the high energy (LAT) photons, first observed in
  short GRBs; and (IV) new
  methods to estimate the Lorentz factor of the outflow.
\end{abstract}

%\maketitle must follow title, authors, abstract
\maketitle

\thispagestyle{fancy}

% body of paper here - Use proper section commands
% References should be done using the \cite, \ref, and \label commands
% Put \label in argument of \section for cross-referencing
%\section{\label{}}

\section{Introduction}

The data provided by {\it Fermi} on GRBs in the past three years since
its launch, are extreme in richness and quality. {\it Fermi} new
capabilities had focused attention on the prompt and early afterglow
emission phases in GRBs, which show rich phenomenology. Both the
quality and quantity of the data enable, for the first time,
systematic study of these early emission phases.

Study of these phases is very important due to its very wide
applicability: during these early phases, one studies the latest
stages of the collapse and the earliest stages of the jet formation -
before self-similar motion wipes the initial outflow conditions.  This
is also what makes this study so challenging: as no two GRBs are
similar, it is very difficult to draw firm conclusions which are
true to {\it all} GRBs.

Confronting {\it Fermi} data with theoretical models show the deficit
of the latter. The data provided by {\it Fermi} did not fit into any
of the existing theoretical models; in some cases, it was in
contradiction to the theoretical expectations. This fact motivated new
ideas, which are continuously being raised and discussed. With this
respect, the contribution of {\it Fermi} is enormous - questions which
were thought to be solved are proven not to be.

In this review, I discuss some of the key {\it Fermi} results, from a
theoretician's perspective. I start by stating in \S\ref{sec:2} what
we are certain about, and then in \S\ref{sec:3} what are the basic
theoretical questions which need to be answered. In \S\ref{sec:4} I
highlight some of {\it Fermi's} unexpected results, and their
influence on our understanding of GRBs.  I then describe in
\S\ref{sec:5} the recent theoretical progress motivated by these
results, and try to point towards additional observational signatures
that could help resolve basic theoretical problems.

\section{GRBs: basic, unquestionable facts}
\label{sec:2}

The field of GRBs is characterized by many uncertainties and a huge
divergence within the GRB population, both in spectral and time
domains.  Nonetheless, after two decades of extensive research there
are few basic, unquestionable facts, which are common to all GRBs and
should be addressed by any theoretical model. It is firm today that
GRBs are:
\begin{enumerate}
\item Transient in nature: no repetition was ever found. The duration
  of the prompt phase vary a lot from burst to burst, and can last
  between a fraction of a second to hundreds of seconds. During this
  phase, the light curve is highly variable.

\item Extragalactic objects, originating at cosmological
  distances.

\item Very energetic, releasing (isotropically equivalent) energy of
  $\sim 10^{49} - 10^{55}$~erg in $\gamma$-rays alone.

\item The observed spectrum is non-thermal. In the vast majority of
  bursts, it has a broken power law shape (the ``Band'' function,
  named after the late David Band), peaking at sub-MeV, with a fairly
  sharp decline at higher energies (see Figure \ref{fig:1}). In a
  small fraction, about $\sim 5\%$, photons were seen up to very high
  energies, $\sim 30 \GeV$. In few bursts there is an additional, high
  energy component which is not related to the original ``Band''
  function \cite{Zhang+11}.

\item Relativistic expansion: very high Lorentz factor, $\Gamma\sim
  10^2 - 10^3$ is required by observations of high energy
  photons. This has solid confirmation by the existence of afterglow
  emission, which follows the interaction of the relativistic ejecta
  with the ambient medium.

\item There are two populations of bursts, separated by their duration
  and hardness: the ``short/hard'' and the ``long/soft''
  \citep{Kouveliotou+93}.  There are firm evidence that long GRBs are
  connected to supernovae \cite[e.g.,][]{Hjorth+03}. Indirect evidence
  suggest that short GRBs originate from binary mergers
  \citep{Berger+05}, but no conclusive evidence yet
  \citep[e.g.,][]{NFP09, Virgili+11}.

\end{enumerate}

\section{Open questions prior to the launch of Fermi}
\label{sec:3}

Based on these observational facts, a general framework, the GRB
``fireball'' model\footnote{Interestingly enough, alternative
  scenarios, such as the ``Cannonball'' model, or the ``fireshell''
  model, have similar basic ingredients.} emerged nearly
two decades ago, and still serves as the main theoretical framework.

The basic phases of this model are best described in terms of the {\it
  energy flow}.  The progenitor, which can be either a collapse of a
massive star or merger of binaries, releases gravitational energy
($E_G$).\footnote{Another potentially significant source of energy is
  the spin energy of the central object.} While part (or even most) of
this energy is released in the form of neutrinos or magnetic flux, a
significant part is eventually converted to kinetic energy ($E_k$):
this is the jet formation episode.

At a second stage, part of the jet kinetic energy is being dissipated
(say, fraction $\epsilon_d \leq 1$). Several mechanisms were suggested
for this dissipation, such as internal shocks \cite{RM94} or magnetic
reconnection \cite{Drenkhahn02, DS02}. Uncertain part of the
dissipated energy ($\epsilon_d E_k$) is used to produce population of
energetic, non-thermal electrons, which emit the observed
$\gamma$-rays.  Other parts may be used to generate magnetic field,
accelerate protons or simply heat the plasma.  The remains of the
kinetic energy is gradually released at later times, producing the
observed afterglow.

This general framework has two strong advantages. First, it is
consistent with all past and current observations. Second, the
existence of the afterglow was predicted by this model, and hence its
detection is a strong confirmation.

This model, however, suffers several very serious drawbacks. First,
the model is heuristic in nature, and many of the details of the
physics are missing. For example, the details of the jet formation
which result in the very high Lorentz factor, $\Gamma \sim 10^2 -
10^3$ as compared to AGNs, in which $\Gamma \leq 30$ are not
explained.  Another example relates to the physics of particle
acceleration to non-thermal distribution, which, although inferred by
the observations, is not well understood.  Other parts of the model
have very little predictive power: for example, internal dissipation
is required to explain the variable light curve, however, the model
does not provide any prediction of the dissipation processes, such as
their radii, amount of kinetic energy that is dissipated, etc.

Thus, major theoretical tasks are to ``fill the gaps'' in this basic 
framework. Very broadly, the theoretical efforts are focused on:

\begin{enumerate}
\item Understanding the nature of the progenitor. 

\item Understanding jet launching mechanism, and the role played by
  $\nu$'s, photons and magnetic field in this process.

\item The dynamics of GRB jets: what causes these jets to have such a
  high Lorentz factor, as opposed to jets in other objects, which have
  much lower Lorentz factors ?

\item Jet composition: what is the role played by leptons, hadrons and
  magnetic field ?

\item Understanding the nature of the dissipation mechanism that leads to
  the emission of $\gamma$-rays.  

\item Radiative processes, and physical explanation to the broad band
  spectrum observed.

\end{enumerate}

In addition to these basic GRB physics questions, other questions
relate to the connection between GRBs and other objects of interest,
such as stellar evolution, host galaxies, binary evolution,
gravitational waves, cosmic rays etc. Additional questions relate to
the use of GRBs as probes of basic physics and cosmology: use of GRBs
as standard candles \citep{Amati+02}, providing limits on violation of
Lorentz invariance \citep{Abdo+09c} etc.

In the past few years there were major theoretical efforts aiming at
addressing these questions. As unfortunately I am not able to summarize
all of the recent works within the page limits of the current
proceedings, I will focus on works which were directly influenced by
recent {\em Fermi} results.  Obviously, as we are observing photons, there is no
direct way to determine the answers to the questions outlined above,
but these have to be deduced indirectly.

\section{Fermi key results}
\label{sec:4}

After three years of operation, one can summarize {\it Fermi} key
results as follows.  The detection rate of the GBM detector is $\sim
250$ bursts/year, which is about the expected rate. However, the LAT
detected only $\sim 10$ bursts/year, namely $\approx 5\%$ of GBM
bursts are observed in the LAT energy range ($\sim 40 \MeV - 300
\GeV$; see Omodei's talk). This fraction is significantly lower than
the pre-launch expectations; however, as will be discussed below,
these expectations were based on mathematical extrapolation of lower
energy data, which did not carry strong physical reasoning.  On the
other hand, the fact that $\sim 30 \GeV$ photons were observed can be
translated, via the opacity argument \citep{KP91} into a stringent
constraint on the Lorentz factor of (LAT) bursts, $\Gamma \sim 10^3$,
which is higher than previously thought.

The spectral properties of the vast majority of {\it Fermi} bursts are
very similar to the spectral properties of the {\em CGRO-BATSE}
bursts. Most GBM bursts are well fitted with a ``Band'' function,
peaking at sub-MeV. The low energy spectral index is, on the average,
somewhat harder than that of the {\it BATSE} bursts, yet within the
errors (photon index $\alpha = -0.95\pm 0.23$ vs. $\alpha = -1.00\pm
0.31$ for the bright bursts)\citep{Nava+11}.

\begin{figure*}[t]
%\begin{figure}
\begin{center}
\includegraphics[width=8cm,angle=0,clip=]{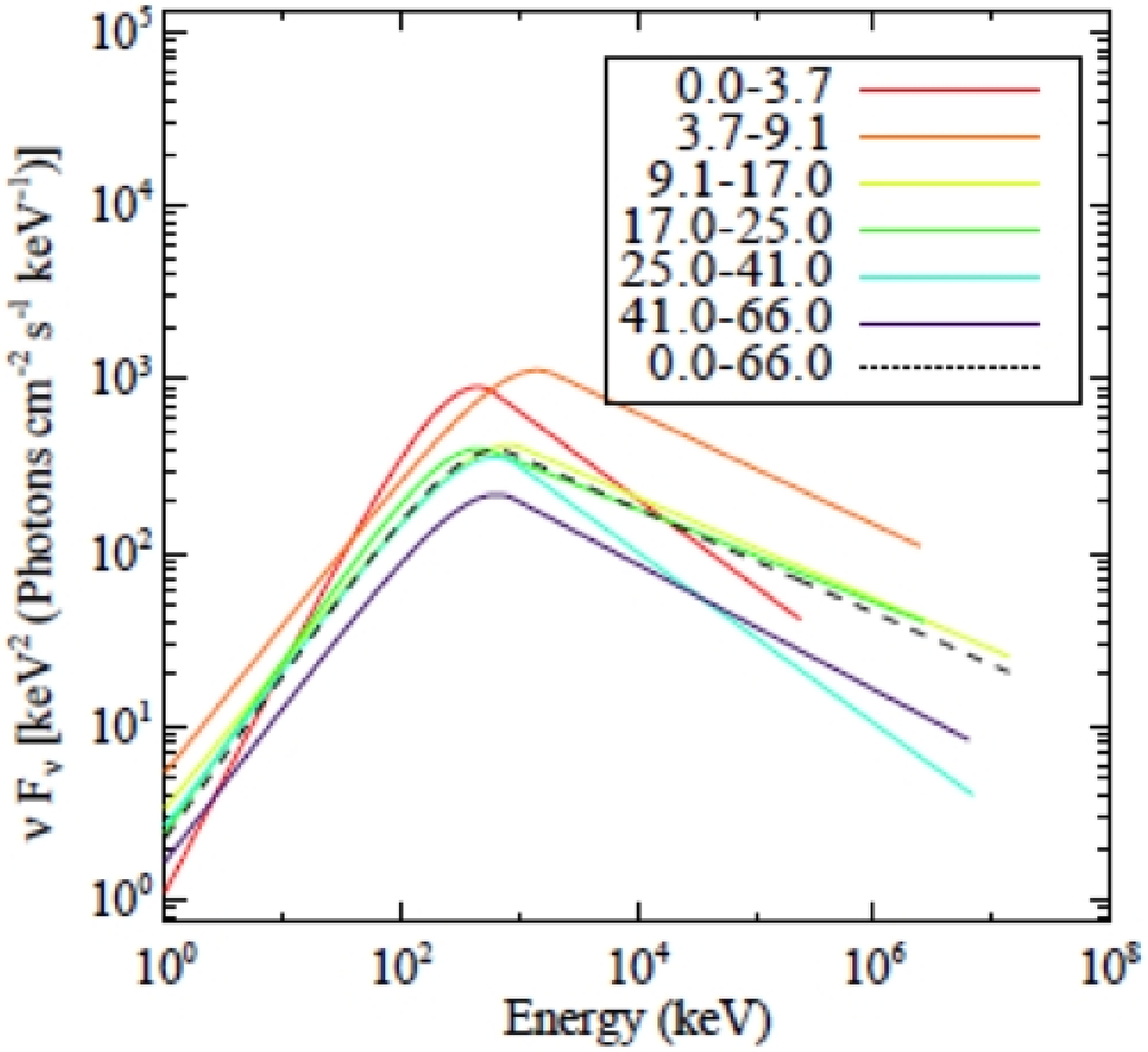}
\includegraphics[width=8cm,angle=0,clip=]{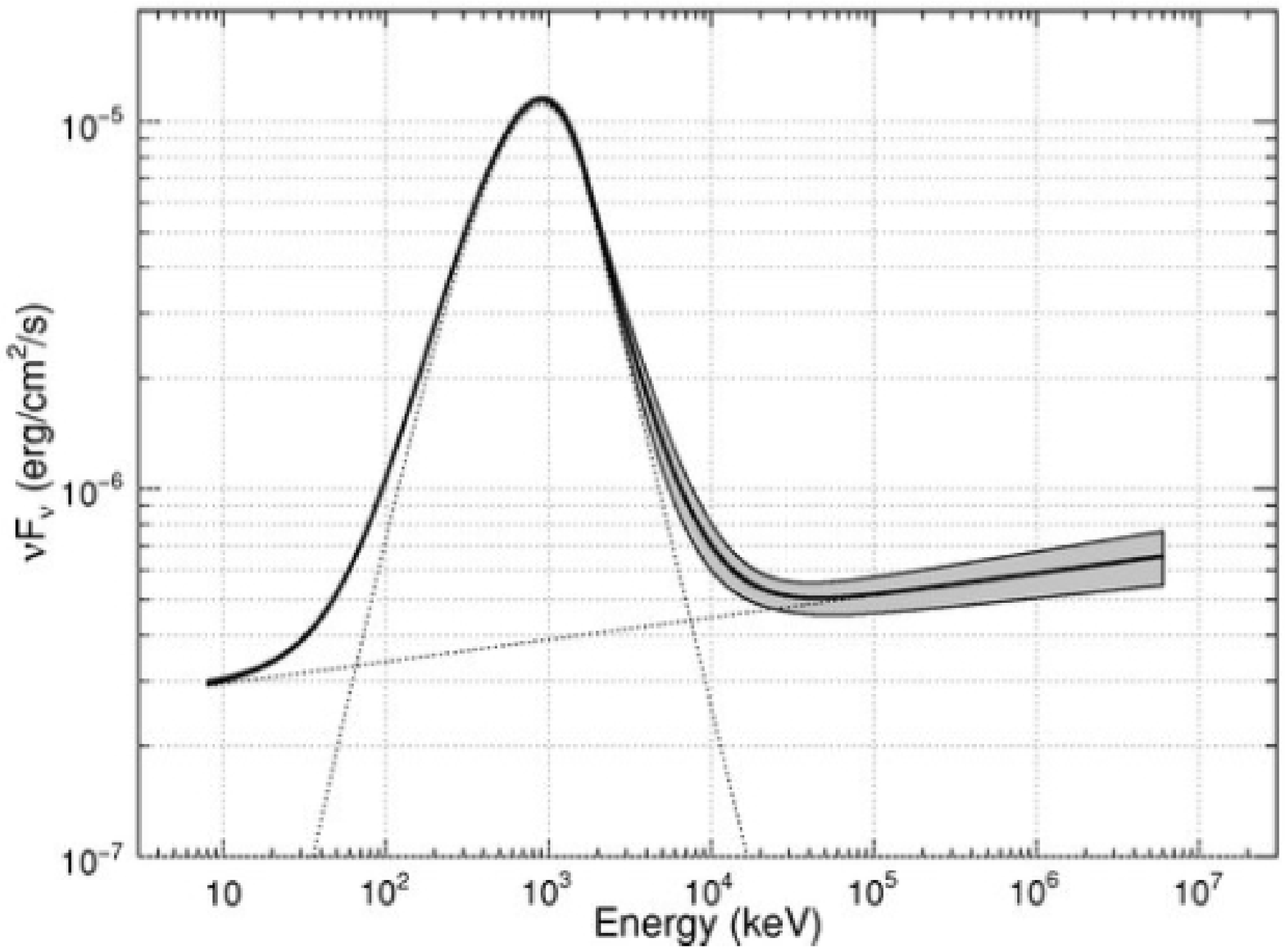}
\end{center}
\caption{Spectra of GRB080916C (left, taken from \citep{Abdo+09b,
    Zhang+11}) at different times
and GRB 090902B (right, taken from \citep{Abdo+09a}) show a clear, ``typical''
broken power law shape, peaking at sub MeV, which is known as the ``Band''
function. There are, however,  qualitative differences: In GRB090902B
there is an extra, high energy power law, which cannot be fitted with
the ``Band'' spectrum. The sub MeV peak of the emission is very hard,
and can best be fitted by a (multi-color) ``Planck'' function
\citep{Ryde+10, Peer+11}. The spectrum of GRB080916C, on the other
hand, is much flatter, and does not require an extra high energy component.}
\label{fig:1}
\end{figure*}

LAT bursts, on the other hand, show a different behavior. Several
bursts show evidence for a separate, extra high energy component, that
is not part of the original ``Band'' fits. This was observed both in long
bursts (e.g., GRB090902B) \citep{Abdo+09a} and in short bursts (e.g.,
GRB090510) \citep{Ackermann+10}. However, this was not observed in the
majority of the LAT bursts \citep{Zhang+11}: for example, the spectrum
of GRB080916C did not show evidence for an extra component
\citep{Abdo+09b}. These qualitative spectral differences are showen in
Figure \ref{fig:1}. It should be stressed that such differences could
not have been seen by {\it BATSE}, due to its limited spectral
coverage, $30 - 2000 \keV$.

\begin{figure}
%\begin{figure}
\begin{center}
\includegraphics[width=8cm,angle=0,clip=]{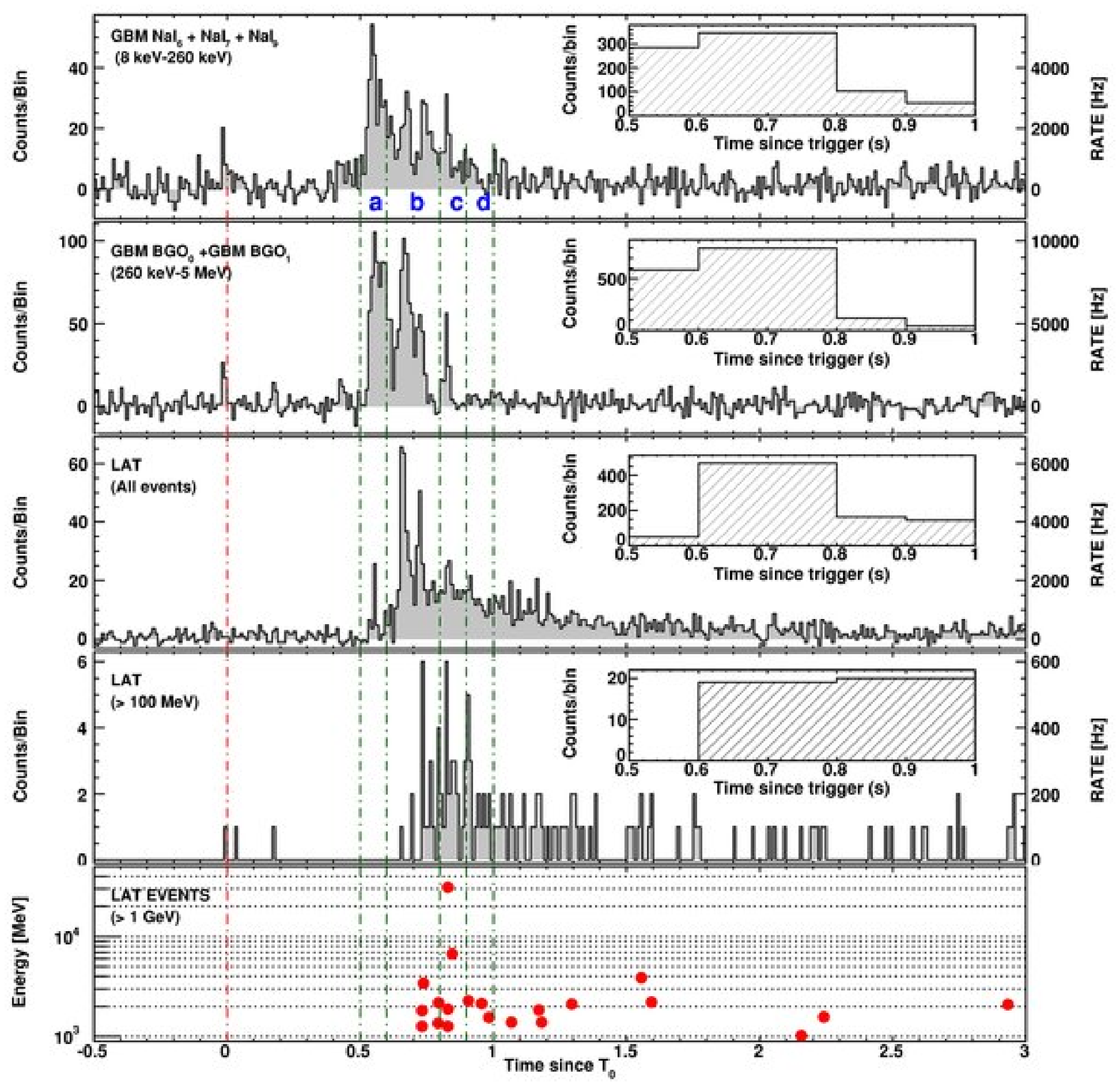}
\end{center}
\caption{
Lightcurve of the short burst GRB090510 \citep{Ackermann+10} show
clear indication for lag in the onset of the high energy (LAT) photons with respect to
the lower energy (GBM) photons, by $\sim$second. Moreover, high energy
emission is long lasting: it extends several seconds after the GBM
emission decayed. Both the delay and the late emission are typical for many LAT bursts.}
\label{fig:2}
\end{figure}

In spite of the fairly low statistics, a clear trend had emerged: in
most (but not all) bursts the high energy (LAT) photons arrive at a
{\it delay} of few seconds with respect to the lower energy (GBM)
photons. This delay is observed in both long and short GRBs: e.g., the
long GRB080916C, and the short GRB090510.  In addition to the delay in
the onset of the high energy photons, another important result is
their extension to late times: LAT photons are observed to be
long-lived. High energy photons are frequently continuously observed
for few seconds after the decay of the low energy (GBM) photons. Both
these features are demonstrated by the temporal behavior of GRB090510
presented in Figure~\ref{fig:2}.

\section{Theoretical implications}
\label{sec:5}

\subsection{Spectral properties: hard spectral slopes}
\label{sec:5.1}

The fact that the low energy spectral index is, on the average,
similar to the low energy spectral index observed by the {\em BATSE},
implies that it is too hard to be accounted for by (optically-thin)
synchrotron emission \citep{Preece+98, GCG03}. This serves as a strong
motivation for alternative scenarios.

Recently, several works considered the effect of inverse-Compton (IC)
scattering on the {\em synchrotron} spectra \citep{Derishev+01, Bosnjak+09,
  Nakar+09, DBD11}. Due to the Klein-Nishina suppression, energetic
electrons are cooled less efficiently than low energy electrons,
resulting in a hardening of the electrons distribution. The resulting
synchrotron emission can be as hard as $F_\nu \propto \nu^0$ under the
appropriate conditions. 

A lot of theoretical attention was given to contribution from the
photosphere, which is arguably the most natural explanation to the
hard slopes observed, being inherent to the ``fireball'' model
\citep{EL00, MR00, DM02, RM05}.  The basic idea is very appealing:
while no combination of synchrotron spectra can produce slopes harder
than $F_\nu \propto \nu^{1/3}$ (in the ``slow cooling'' regime) or
$F_\nu \propto \nu^{-1/2}$ (in the ``fast cooling'' regime) , it is
possible to produce the harder ``Band'' function by broadening Planck
spectra. Thus, in recent years, theoretical research in this field was
focused on mechanisms that can broaden Planck spectrum.

Spectral broadening can be achieved in two ways.  First, energy
dissipation below the photosphere produces a population of non-thermal
electrons, which emit radiation. Such a dissipation can result from
internal shocks, magnetic reconnection or collisional heating. Since
the dissipation is assumed to take place in region of high optical
depth, multiple Compton scattering dominates the spectra, which can be
broader than Planck if the optical depth is not too high. This
scenario gained broad interest recently \citep{PMR05, PMR06,
  Giannios06, Giannios08, Ioka10, LB10, Beloborodov10, Fan10, Toma+11,
  Vurm+11, Bromberg+11, Ryde+11}.  Second, broadening is caused by
contribution of off-axis emission \citep{Peer08, PR11, Beloborodov11,
  Lazzati+11, Mizuta+11}.  While off-axis photons are sub-dominant as
long as the inner engine is active, they broaden the Planck
spectrum. They become dominant once the inner engine decays.
Quantitative results depend on the jet geometry.

On the observational side, following pioneering works by {\it
  Ryde} \citep{Ryde04,Ryde05}, in recent years there are several
successful attempts to identify a thermal (Planck) component in
existing broad-band 
spectra \citep{RP09, McGlynn+09, Ryde+10, Ryde+11, Larsson+11,
  Guiriec+11, Burgess+11}. Such a ``pure'' Planck spectra can be
expected if (I) energy dissipation occurs very deep in the flow or only
outside the photosphere, and (II)  the photospheric radius is
close to the coasting radius, so that adiabatic energy losses are
small. The clear identification of this component, albeit in only a
limited number of bursts, is a strong motivation for continuous
theoretical research.

\subsection{High energy emission:  spectral properties and delayed
  onset}

Understanding the origin of the high energy (LAT) emission is
difficult, because of the confusing results: while in many LAT bursts
(e.g., GRB080916C) the high energy component smoothly connects to the
low energy part, and is part of the ``Band'' function, in others
(e.g., GRB090902B) it is spectrally separated (see Figure \ref{fig:1}). 

Clearly, emission from the photosphere cannot, by itself, explain a
separated spectral component at these energies.  Thus, within the
context of the photospheric emission models, a second episode of
energy dissipation above the photosphere is required in order to
explain $\sim$GeV emission. Inclusion of this additional dissipation
can provide very good fits to the data \citep{Peer+11}.  This scenario
is thus consistent with a separation of the spectral component (as in
GRB09092B), as well as with the observed delay of the high energy
component.  Alternatively, the spectral separation of this component
may result from a different origin: it was suggested by several
authors that while the ``Band'' part has a leptonic origin, the high
energy part may have hadronic origin \citep{Asano+09,Asano+10,
  Razzaque10}, or may originate from an expanding ``Cocoon''
\citep{Toma+09}.

Many LAT GRBs show high energy emission which seems to be smoothly
connected to the low energy part, hinting towards common origin (e.g.,
GRB80916C). The temporal evolution of the decay at these energies,
$F_\nu (t) \propto t^{-1.5}$ is consistent with having external
origin. Thus, it was proposed by several authors \citep{KB09, KB10,
  Ghisellini+10b, Ghirlanda+10, Wang+10, He+11} that the high energy
component results from energy dissipation by the external shock,
similar to the afterglow emission. Thus, according to this view, the
high energy emission is in fact part of the afterglow emission, which
naturally explains the extension of the emission to late times.  On
the other hand, a detailed spectral analysis showed that at least part
of the GeV emission must have internal origin \citep{PN10,
  Maxham+11}. Moreover, the smooth spectral extrapolation to the low
energy part (the Sub MeV peak and below) implies that either the low
energy part also has an external origin, in which case it is difficult
to explain the hardness of the spectral slopes, or that both
components ``conspire'' to smoothly match. Over all, the origin of the
GeV emission is still uncertain, and more data is needed.

\subsection{Jet Composition and dynamics}

As photospheric emission is an inherent part of the classical,
baryonic ``fireball'' model, the fact that it is not observed in most
bursts challenges this model.  The lack of observed photospheric
signal in GRB080916C was used to argue in favor of
magnetically-dominated outflow \citep{ZP09}.

This idea has a firm theoretical basis. Significant developments in
numerical GRMHD codes in the past few years enabled a detailed study
of jet launching \citep{Tchekhovskoy+08,Narayan+10, Rezzolla+11}.
These works mark the first steps towards realization of the
\citet{BZ77} and \citet{ BP82} mechanisms, in which the magnetic field
plays a crucial role as energy mediator in jet production. Thus, if
indeed these are the jet launching mechanisms that work in nature, one
expects highly magnetized outflow, $\sigma \equiv u_B / \rho c^2 >>1$
close to the jet-launching site. Here, $u_B$ is the (electro-)
magnetic energy density and $\rho$ is the baryon density.

Motivated by these results, in recent years there were numerous
attempts to study the properties of magnetized outflows. While studies
in this field are not new and were done prior to {\it Fermi} era
(e.g., \citep{Thompson94, DS02, LB03}), {\em Fermi's } recent results
stimulated a renewed interest.  Research in this field is currently
focused on bridging the theoretical gaps in all aspects of magnetized
jet physics: (I) creation of relativistic jets in magnetars
\citep{Bucciantini+09, Giannios10, Metzger+11, Granot+11}; (II)
dynamics, energy dissipation and efficiency of magnetic reconnection
and magnetized shock waves \citep{MA10, Lyubarski10, Komissarov+10,
  MU11, ZY11, Narayan+11}; (III) particle acceleration in magnetized
outflows \citep{SS09, SS11}, and (IV) the resulting radiative
signature \citep{Giannios08, GS09, MR11, ZY11}. While the current
picture is far from being complete, this is a very active research
field.

Finally, detection of high energy photons imply, using the opacity
argument, high Lorentz factor, $\Gamma \simeq 10^3$ in few {LAT}
bursts \citep{Racusin+11, Ioka10}.  These values are more than an
order of magnitude higher than the typical Lorentz factor in AGNs,
$\Gamma \leq 30$, and are not theoretically understood. Several works
in recent years were focused on finding new methods to determine the
Lorentz factor \citep{Peer+07}, and to find better constraints on the
Lorentz factor of the outflow. Indeed, a more advanced analysis using
the assumption of multi-zone emission, or taking geometrical
corrections showed that the Lorentz factor may not be as high, and may
be limited to few hundreds \citep{Zou+11, Hascoet+11}.

\section{Summary}
\label{sec:6}

{\em Fermi} results focused attention on the {\em prompt} emission
phase in GRBs, which is rich in phenomenology, and is poorly
understood. While the general ``fireball'' framework that was
constructed during the 90's still holds, a lot of gaps in the
theoretical understanding still exist.

In recent years significant theoretical progress was made in
understanding the physics of the prompt emission. I highlighted here a
few active research areas in which, to my opinion, {\em Fermi} results
were very influential.
\begin{enumerate}
\item{\em Progenitor, jet launching and composition}.
A renewed interest in magnetized models
  and magnetars as GRB progenitors had emerged, largely stimulated by
  recent {\em Fermi} results. Research is focused on all aspects of
 (relativistic)  magnetized outflows, from jet launch, dynamics to the
 observed  signal.

\item{\em Radiative processes and spectral properties}. A lot of
  effort was given to understanding the role played by photospheric
  emission, and in particular mechanisms that can be used to broaden
  Planck spectrum, so that it will resemble the observed ``Band''
  spectra. Effort is given to understanding signatures of magnetized
  outflows. 

\item{\em Jet dynamics and origin of high energy emission}. In spite
  of numerous efforts, the origin of high energy emission is still
  unclear. This is because of the confusing results, which show in
  many bursts smooth spectral extrapolation between the high energy
  and lower energy photons, while a separate component in other
  bursts.

\item{\em Accurate measurement of the Lorentz factor}. The high values
  of $\Gamma \sim 10^3$ inferred by the existence of high energy
  photons are challenging. They motivated a more careful analysis of
  the data, which indeed show somewhat lower values. Still, no
  theoretical understanding of these values exists.
  
\end{enumerate}

\bigskip % extra skip inserted
\begin{acknowledgments}
I wish to thank {\it Felix Ryde}, {\em Bing Zhang} and {\em Raffaella
  Margutti} for many useful discussions
and comments.

\end{acknowledgments}

\bigskip % extra skip inserted
% Create the reference section using BibTeX:
%\bibliography{basename of .bib file}

%\bibliography{/Users/apeer/Documents/Bib/abbrevs,/Users/apeer/Documents/Bib/short_abbrevs,/Users/apeer/Documents/Bib/bib_apeer}

%\begin{thebibliography}{9}   % Use for  1-9  references

\end{document}